\documentclass[showpacs,superscriptaddress,amsfonts,amsmath,floatfix,twocolumn]{revtex4-1}

\usepackage{graphicx}
\usepackage{hyperref} 
\usepackage{verbatim} 
\usepackage{url}
\usepackage{epstopdf}
\usepackage[normalem]{ulem}
\usepackage{color}
\usepackage{ulem}
\usepackage{dsfont}
\usepackage[utf8]{inputenc}
\DeclareUnicodeCharacter{2010}{-}
\usepackage{float}
\usepackage{amsmath}
\usepackage{comment}
\usepackage{breqn}
\usepackage{amssymb}
\usepackage{bbold}

\DeclareMathOperator{\Tr}{Tr}

\renewcommand\vec[1]{\ensuremath\boldsymbol{#1}}

\newcommand{\affA}{Abdus Salam International Centre for Theoretical Physics, Strada Costiera 11, 34151 Trieste, Italy}
\newcommand{\affB}{Department of Physics and Astronomy, University of Florence, 50019 Sesto Fiorentino, Italy}
\newcommand{\affC}{INO-CNR Istituto Nazionale di Ottica del CNR, Sezione di Sesto Fiorentino, 50019 Sesto Fiorentino, Italy}
\newcommand{\affD}{LENS, European Laboratory for Non-Linear Spectroscopy, 50019 Sesto Fiorentino, Italy}

\begin{document}
\title{Engineering entanglement Hamiltonians with \\strongly interacting cold atoms in optical traps}

\author{R. E. Barfknecht}
\email{barfknecht@lens.unifi.it}
\affiliation{\affC}
\affiliation{\affD}

\author{T. Mendes-Santos}
\affiliation{\affA}

\author{L. Fallani}
\affiliation{\affB}
\affiliation{\affD}
\affiliation{\affC}

\date{\today}
\begin{abstract} 
\noindent We present a proposal for the realization of entanglement Hamiltonians in one-dimensional critical spin systems with strongly interacting cold atoms. Our approach is based on the notion that the entanglement spectrum of such systems can be realized with a physical Hamiltonian containing a set of position-dependent couplings. We focus on reproducing the universal ratios of the entanglement spectrum for systems in two different geometries: a harmonic trap, which corresponds to a partition embedded in an infinite system, and a linear potential, which reproduces the properties of a half-partition with open boundary conditions. Our results demonstrate the possibility of measuring the entanglement spectra of the Heisenberg and XX models in a realistic cold-atom experimental setting by simply using gravity and standard trapping techniques.
\end{abstract}


\maketitle

\paragraph*{Introduction.} The study of entanglement in quantum many-body systems \cite{RevModPhys.80.517,RevModPhys.81.865} has become one of the major efforts in the physics community, not only because it is a central feature of quantum theories, but also due to its potential for describing quantum phases of matter and topological order \cite{Calabrese_2009,PhysRevB.81.064439,PhysRevLett.101.010504,PhysRevLett.104.130502,PhysRevLett.104.180502,PhysRevLett.108.196402,PhysRevLett.109.237208}. Directly measuring entanglement in experiments, on the other hand, has proven to be a challenging task. This is due, in particular, to the difficulty of obtaining the full-state tomography \cite{PhysRevLett.120.025301} of many-body systems. Nevertheless, outstanding progress has been made in recent years with respect to the extraction of the entanglement properties of quantum systems, both in terms of theoretical proposals \cite{PhysRevLett.106.150404,PhysRevLett.93.110501,PhysRevLett.109.020505,PhysRevLett.109.020504,PhysRevLett.120.050406,PhysRevA.97.023604,Morera2019} and experiments \cite{Islam2015,PhysRevLett.115.035302,Hauke2016,Brydges260}. In common, these works have the feature of employing indirect measurement protocols, either through the probing of correlations or by interference of identical copies of the system. 

In this context, it is highly desirable to have at hand alternative proposals \cite{PhysRevX.6.041033} for the measurement of entanglement that are at the same time direct - in terms of quantities which are ordinarily accessible in experiments - and scalable.
A significant step in this direction has been taken recently \cite{Cardy_2016,Dalmonte2018,PhysRevB.98.134403,Mendes_Santos_2020}, with works showing that the entanglement spectrum \cite{PhysRevA.78.032329,PhysRevLett.113.060501} of lattice Hamiltonians can be reproduced by obtaining the physical spectrum of a Hamiltonian with the same general properties, but with a set of spatially varying coupling parameters. This notion is based on the Bisognano-Wichmann (BW) theorem, which originally describes entanglement Hamiltonians for continuous systems in the context of quantum field theory \cite{bw1,bw2}. Later developments have shown \cite{PhysRevB.100.155122} that particular features of the original model, such as boundary conditions, have a direct effect on the functional form of the couplings in the physical Hamiltonian. Such a relation hints at the prospect of reproducing the entanglement properties of discrete systems with Hamiltonians with specifically designed non-homogeneous couplings. 

An immediate possibility that arises in this context is the simulation of entanglement Hamiltonians with trapped systems of cold atoms \cite{bloch1}, where optical confinement, atomic interactions and internal states can be manipulated with remarkable precision. Such setups are therefore ideal candidates for quantum simulations of condensed matter systems. \cite{doi:10.1080/00018730701223200,Bloch2008,Bloch2012,Gross995}. On the theoretical side, it has been demonstrated that if the interactions between atoms are strong enough, the system can be mapped from the continuum to a spin chain, where nearest-neighbor couplings are determined by the local trapping geometry \cite{artem1,deuretz2,pu1,massignan}. Experimentally, the strongly correlated regime is accessible with cold atoms both for fermionic \cite{PhysRevLett.94.210401,Liao2010,jochim1,pagano,Hilker484} and bosonic \cite{paredes,weiss2,trotzky,haller1,haller2} gases. In fact, the validity of the theoretical approach described above has been verified in recent experiments \cite{jochim3}.  

Inspired by this perspective, in the present work we show how to engineer the entanglement Hamiltonians of spin systems, such as the Heisenberg (XXX) and XX models, by considering a strongly interacting two-component system of cold atoms in effectively one-dimensional optical traps. Our first application is the case of harmonically trapped atoms, which gives rise to a spin chain where the couplings approximately follow a parabolic distribution. As we will show, this remarkably ordinary assumption is enough to reproduce the universal ratios of the entanglement spectrum of a partition embedded in an infinite system with periodic boundary conditions. The second application assumes the presence of a linear potential, which in turn results in a set of linearly increasing couplings for the spin chain. In this regime, the physical Hamiltonian can reproduce the ratios of the entanglement spectrum of a half-partition of a spin system with open boundary conditions. The simple linear form of the potential additionally poses the intriguing possibility of using gravity as a tool for probing entanglement. 
To account for other experimental-related aspects, we also include Density Matrix Renormalization Group (DMRG) simulations of the continuum away from the strongly interacting limit, as well as the expected effect of finite temperatures on measurable quantities such as the dynamical structure factor.

\paragraph*{System description}\label{description}

We calculate the entanglement properties of one-dimensional spin chains such as the Heisenberg model, $H=J\sum_{i=1}^{N-1}\vec{\sigma}_i\cdot \vec{\sigma}_{i+1}$, and the XX model $H=J\sum_{i=1}^{N-1}\sigma_i^x \sigma_{i+1}^x+\sigma_i^y \sigma_{i+1}^y$,
where $\vec{\sigma}$ denotes the Pauli vector and $J$ is a homogeneous coupling. After finding the ground state solution $\vert\Psi\rangle$ for one of these models, the reduced density matrix for a subsystem $A$ can be calculated by tracing over the remaining subsystem $B$ with $\rho_A=\Tr_B\vert \Psi \rangle \langle \Psi \vert=e^{-H_A}/Z_A$, where $H_A$ is called the {\it entanglement Hamiltonian} and $Z_A$ is a normalization constant. The entanglement spectrum for partition $A$ is then obtained as the set of eigenvalues $\{\epsilon_n\}=-\ln\{\epsilon^{A}_n\},$ where $\epsilon_n^{A}$ denotes the eigenvalues of the reduced density matrix. 

\begin{figure}
\centering
\includegraphics[width=0.45\textwidth]{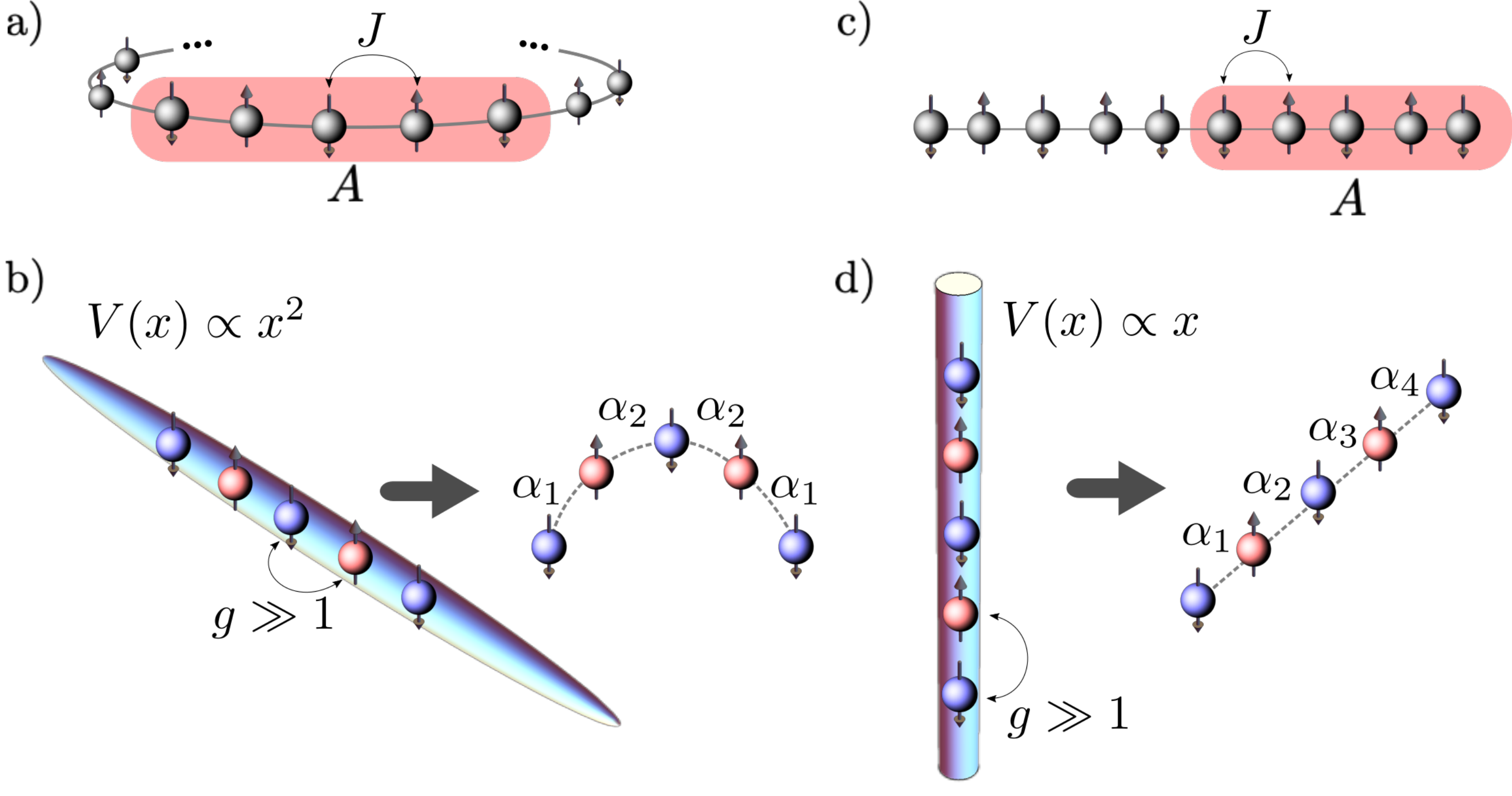}
\caption{The entanglement properties of one-dimensional spin chains can be simulated with strongly interacting systems of cold atoms in optical tubes. These systems can be mapped to spin chains where the couplings are determined by the local trapping geometry.
a) A partition embedded in an infinite spin system with periodic boundary conditions ($T_1$), can be simulated with b) a strongly interacting ($g\gg 1$) two-component system of cold atoms in a harmonic trap. We model this system as a spin chain where the exchange coefficients $\alpha_i$ follow the distribution of an inverted parabola. c) A half-partition of a finite spin chain with open boundary conditions is reproduced by d) cold atoms in a linear potential, which can be realized by gravity. This system is mapped into a spin chain where the coefficients increase linearly with the position.}
\label{fig1}
\end{figure}

The BW theorem for discrete systems \cite{Dalmonte2018} states that the entanglement Hamiltonian $H_A$ can be equivalently calculated as
\begin{equation}\label{BWdiscrete}
    H_A\propto \sum_i J_i \vec{\sigma}_i\cdot \vec{\sigma}_{i+1},
\end{equation}
where $J_i$ is a set of position-dependent couplings. Remarkably, predictions from conformal field theories predict \cite{Casini2011,PhysRevB.100.155122} that these couplings should be given by
\begin{equation}\label{exactbc}
J_i\propto \frac{i(N-i)}{N}\,\,\,\,\,\,\text{and}\,\,\,\,\,\,
J_i\propto i,
\end{equation}
for entanglement Hamiltonians corresponding to partitions embedded in an infinite system with periodic boundary conditions (which we label $T_1$) and half-partitions in systems with open boundary conditions ($T_2$), respectively (see Fig. \ref{fig1}). These results indicate that the entanglement Hamiltonians of spin systems can be simulated by constructing a {\it physical} Hamiltonian with a set of properly engineered couplings. A suitable platform for such an endeavor is an effectively one-dimensional cold atomic gas with a Hamiltonian described by
\begin{equation}\label{h1}
H=\sum_{i=1}^{N}\left(-\frac{1}{2}\frac{\partial^2}{\partial x_i^2}+V(x)\right)+ g\sum_{i}^{N_\uparrow}\sum_j^{N_\downarrow}\delta(x_{i,\uparrow}-x_{j,\downarrow}),
\end{equation}
where the first term on the right-hand side includes the presence of a trapping potential $V(x)$ and the second term accounts for contact interactions between atoms in different internal states (we assume a two-component gas where the internal states are labeled as $\uparrow,\downarrow$). The strength of the interactions is set by the parameter $g$ (in units of $\hbar^2/ml$ where $l$ is the characteristic length of the system), and we initially consider the fermionic case where interactions between atoms in the same internal state are forbidden by the Pauli principle (we will relax this restriction later when discussing the simulations of the XX model).

In the limit of strong repulsion ($g\gg 1$), the wave function of a system described by Hamiltonian \eqref{h1} is given by $\Psi=\sum_k a_k P_k\Phi_0$, where $\Phi_0=\Phi_0(x_{\uparrow 1},...,x_{\uparrow N_{\uparrow}},x_{\downarrow 1},...,x_{\downarrow N_{\downarrow}})$ is the wave function in the limit of infinite repulsion and each term of the sum includes a permutation $P_k$ (with amplitude $a_k$) of the coordinates. By employing the Hellmann-Feynman theorem $\frac{dE}{dg}=\langle \Psi \vert \frac{dH}{dg} \vert \Psi \rangle$ along with boundary conditions for the contact interactions $\left(\frac{\partial\Psi}{\partial x_{\uparrow}}-\frac{\partial\Psi}{\partial x_{\downarrow}}\right)\rvert_{x_\uparrow-x_\downarrow=0^-}^{x_\uparrow-x_\downarrow=0^+}=2g \Psi(x_{\uparrow}=x_{\downarrow})$, Eq.~\eqref{h1} can be mapped into the following spin chain \cite{deuretz2,artem2,Barfknecht2019} (see Supplemental Materials \cite{sm} for details)

\begin{equation}\label{spinchain}
H=E_0-\frac{1}{2}\sum_{i=1}^{N-1}\frac{\alpha_{i}}{g}(1-\vec{\sigma}_i\cdot \vec{\sigma}_{i+1}),
\end{equation}
where $E_0$ denotes the energy of the system at the fermionization limit where $g=\infty$. This mapping can be interpreted as a perturbation with respect to the limit of infinite repulsion (notice that we can rewrite this Hamiltonian in terms of the permutation operator $\Pi_{i,i+1}=\frac{1}{2}(1+\vec{\sigma}_i\cdot \vec{\sigma}_{i+1})$ \cite{deuretz2}). By diagonalizing this Hamiltonian we find the amplitudes $a_k$ for the wave function $\Psi$ and the energy spectrum in the limit of strong interactions. Another fundamental aspect of the approach described above is that, in Eq.~\eqref{spinchain}, the position-dependent exchange coefficients $\alpha_i$ are obtained from the properties of the spatial wave function $\Phi_0$ using the following relation 
\begin{equation}\label{geo}
\alpha_{i}=\int_{x_1<...<x_N-1} dx_1\,...\,dx_{N-1}\Big|\frac{\partial \Phi_0}{\partial x_N}\Big|^2_{x_N=x_i},
\end{equation}
which is independent of spin. The many-body wave function described by $\Phi_0$ is constructed as the Slater determinant of the $N$ lowest single-particle states in the trapping potential $V(x)$. Aside from the symmetry considerations regarding permutations of particles, this wave function is the same as the one that describes a Bose gas in the limit of infinite repulsion \cite{girardeau}.
In Eq.~\eqref{spinchain}, the energy $E_0$ is thus simply given by the sum of the energies of each of these single-particle states. It is therefore clear that once we determine the geometry of the trap $V(x)$, we can obtain the exchange coefficients that describe it in Eq.~\eqref{spinchain}. While solving the integrals given by Eq.~\eqref{geo} can be cumbersome for large $N$, efficient methods which exploit the determinant properties of $\Phi_0$ are available \cite{conan,deuretz_mdist}. In Fig.~\ref{fig1} we show a sketch of the protocol adopted as a proposal for the simulation of entanglement with trapped systems of cold atoms.

We have thus established our proposal for simulating the entanglement spectrum of a spin chain with a two-component atomic system, where the energy levels can be obtained by standard spectroscopy techniques \cite{PhysRevLett.101.250403}. The presence of strong interactions and external trapping potentials not only allows for benchmarking the results against the limit of infinite repulsion, but also automatically generates the set of coupling needed for engineering entanglement Hamiltonians. The relation between entanglement and inhomogeneities in the underlying geometry of the system has been also explored, particularly for non-interacting fermions, in \cite{PhysRevA.97.023605,SciPostPhys.2.1.002,Tonni_2018,Murciano_2019}. 

\paragraph*{Results.}

\begin{figure}
\centering
\includegraphics[width=0.45\textwidth]{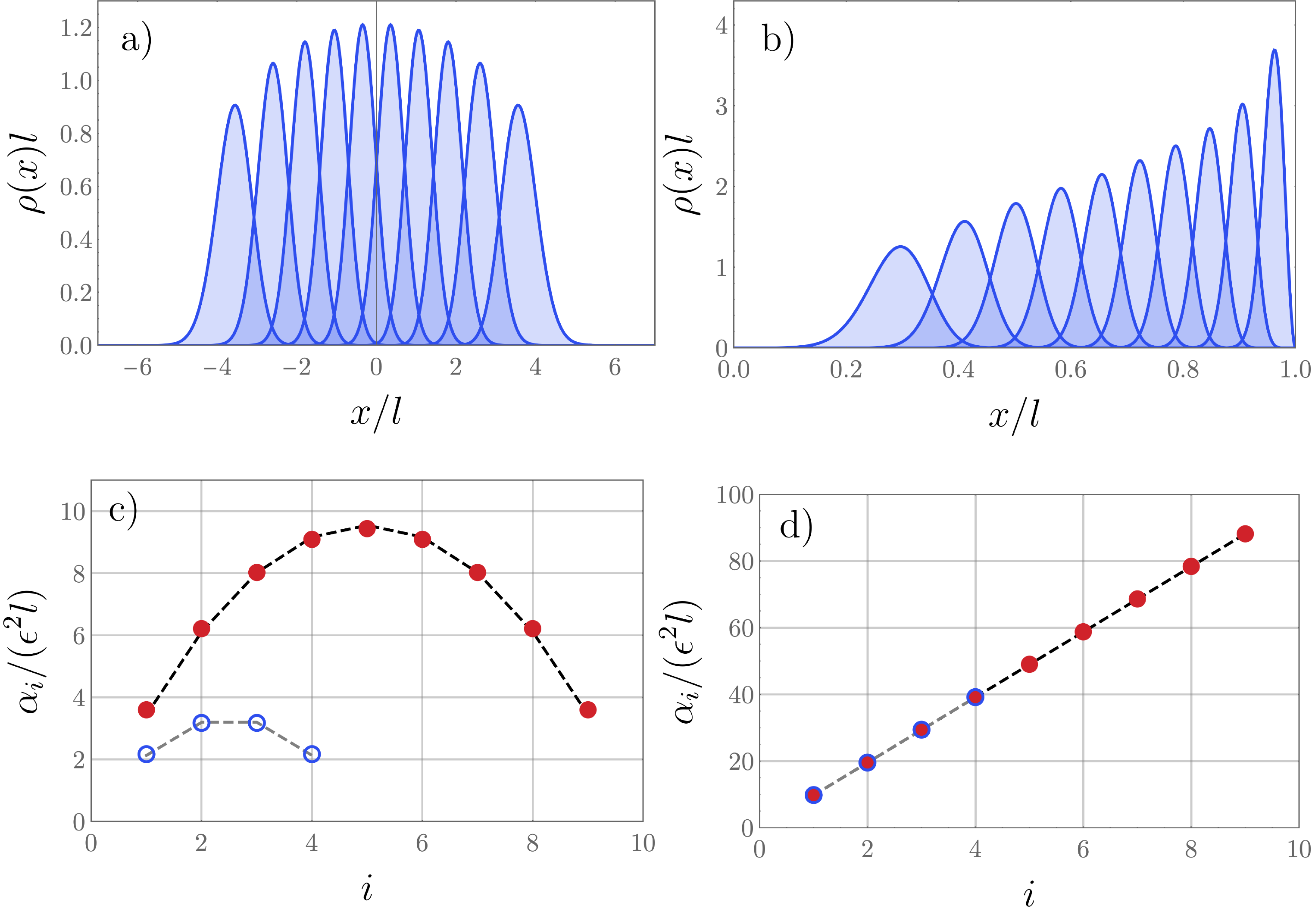}
\caption{Upper row: spatial distributions for $N=10$ spinless fermions in the cases of a) the harmonic trap and b) a finite system of length $l$ with a linear potential. The total density in each case is normalized to the total number of particles $N$. Bottom row: numerical values of the exchange coefficients for the c) harmonic trap and d) the linear trap. The red dots indicate the results for $N=10$, while the blue circles show results for $N=5$. The black and gray dashed curves display results for the fits obtained with Eq.~\eqref{exactbc}.}
\label{fig2}
\end{figure}

We now calculate the properties of the atomic systems described above, including the particle distributions and numerical values of the exchange coefficients for a given trapping potential. In Fig.~\ref{fig2}, we show the single-particle densities for the wave function $\Phi_0$ for spinless fermions in a) the harmonic trap $V(x)=x^2/2$ and in b) the case of a finite system confined by hard walls at $x=0$ and $x=l$ and exposed to a linear potential $V(x) = V_0 (l-x)$. Here we observe how the overlaps between neighboring particles are affected by the underlying geometry. This feature is reflected in the numerical values of the exchange coefficients $\alpha_i$, which are shown in Fig.~\ref{fig2} c) and d). The distribution of these values, for the harmonic trap, is symmetric across the origin (therefore it is enough to calculate at most the first $N/2$ coefficients) and has the shape of an inverted parabola \cite{Levinsene1500197,pu2}. For the tilted potential, we find that the coefficients increase linearly and are simply given by $\alpha_i=V_0 i$. In both Figs.~\ref{fig2} c) and d), we additionally show a comparison with the results obtained by fitting the functions in Eq. \eqref{exactbc}, for two different sizes of partitions. We find excellent agreement between the set of couplings given by Eqs. \eqref{exactbc} and those generated by the two choices of trapping potential.

We are now able to calculate the exact entanglement spectrum for these different types of partitions and compare to results obtained with a physical Hamiltonian with spatially varying couplings. For the Heisenberg model, the physical Hamiltonian is naturally given by Eq. \eqref{spinchain}. The XX model can analogously by reproduced by initially considering a bosonic system in the continuum, where interaction between atoms in the same internal states are allowed. This results in an additional term in Eq. \eqref{spinchain} given by $-\frac{1}{2}\sum_{i=1}^{N-1}\frac{\alpha_i}{\kappa g}\left(1+\sigma_i^z \sigma_{i+1}^z\right)$. For a case of imbalanced interactions ($\kappa=2$) this bosonic system maps into the XX model \cite{artem2}. 

\begin{figure}
\centering
\includegraphics[width=0.45\textwidth]{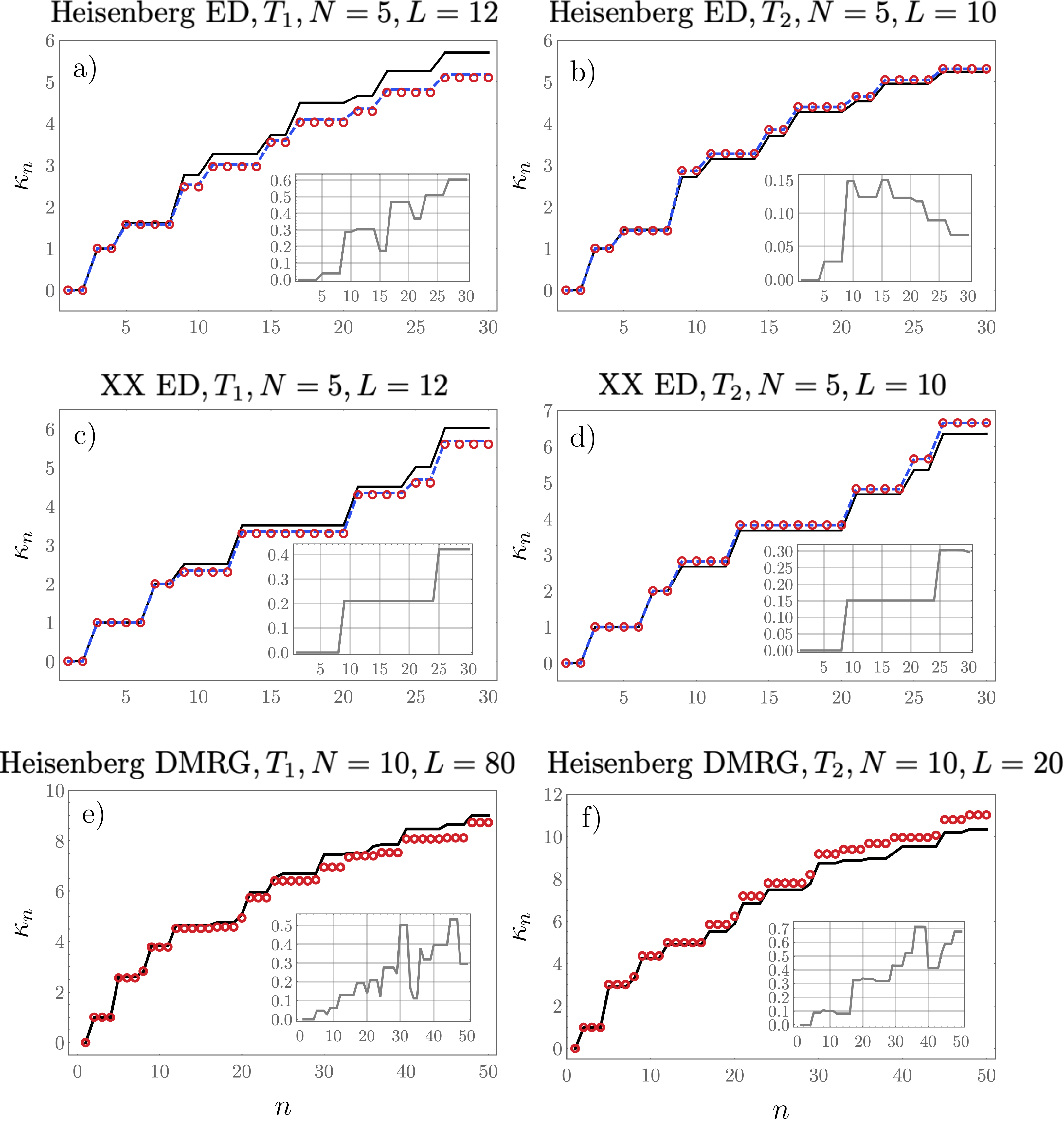}
\caption{Comparison of the universal ratios in Eq. \eqref{ratios} for the a)-b) Heisenberg and c)-d) XX models with $N=5$. The left column show the results for partitions $T_1$ (black solid lines, total size of the system $L=12$), compared to the case of harmonically trapped atoms (red circles). The right column show the analog results for partitions $T_2$ (total size of the system $L=10$) and atoms in a linear potential. In the first four panels we additionally include the results for a Hamiltonian with $N=5$ and a set of ideal couplings given by Eq. \eqref{exactbc} (blue dashed lines). In e)-f) we present the results for the Heisenberg model with larger size ($N=10$). The results for the partitions are obtained with DMRG for systems of total size e) $L=80$ and $L=20$, respectively. The interaction strength for the trapped atoms is set as $g=20$. In all plots, the insets show the absolute numerical difference between the red circles and the black lines.}
\label{fig3}
\end{figure}

Our focus is on obtaining universal ratios for the entanglement spectrum, defined as
\begin{equation}\label{ratios}
\kappa_n=\Big|\frac{\epsilon_n-\epsilon_0}{\epsilon_r-\epsilon_0}\Big|,
\end{equation}
where $n$ denotes the energy level, $\epsilon_0$ is the ground state energy and $\epsilon_r$ is a reference energy level (unless stated otherwise, we fix $r=3$ - the second excited state). In Fig. \ref{fig3} a)-d) we show the results for this quantity obtained by exactly diagonalizing different systems. The physical spectrum of the harmonically trapped strongly interacting Hamiltonian is compared to a partition embedded in a system with periodic boundary conditions, while the case of linear potential is compared to a half-partition of the same size in a system with open boundary conditions. Additionally, we include the results for an ``ideal" physical Hamiltonian with a set of couplings provided by Eq. \eqref{exactbc}. The agreement between the results obtained with three different approaches is particularly remarkable for small values of $n$, corresponding to the low-energy part of the entanglement spectrum. The discrepancy with the exact results obtained from the reduced density matrix (black solid curves) at higher energies stems mainly from finite-size effects.

In Fig. \ref{fig3} e)-f), we include also results obtained with DMRG, where we see that the agreement between the entanglement spectrum of homogeneous partitions of the Heisenberg model and the physical spectrum of trapped fermions also holds for larger systems.

\paragraph*{Experimental details.} Effectively one-dimensional systems of cold atoms are currently realized in the lab by loading the atoms into tight optical waveguides, where the confinement along the transverse direction is much larger than the longitudinal one. These waveguides can be provided by optical lattices, where the transverse confinement could easily reach values on the order of $\omega_\perp=2\pi 50$ kHz, with $\hbar \omega_\perp$ much larger than temperature and atomic chemical potential.

Different spin states can be simulated by exploring the internal degrees of freedom of particular atomic species. Usually, these are hyperfine states in alkali elements (such as fermionic ${}^6$Li and bosonic ${}^{87}$Rb), but nuclear-spin states in alkaline-earth elements like ${}^{173}$Yb or ${}^{87}$Sr can also be used. These last cases also provide the opportunity of exploring additional internal states with SU$(N)$ symmetry) \cite{pagano,mancini}. 

With state-of-the-art atom trapping techniques there are ample possibilities of controlling the potential parameters to engineer the required exchange coupling $\alpha_i$. Typical axial harmonic trapping frequencies for the protocol shown in Fig. \ref{fig1} b) lie in the range $\omega_T = 2\pi (10 - 10^3)$ Hz. The protocol shown in Fig. \ref{fig1} d) can be realized either by means of magnetic/optical gradients or by exploiting the effect of gravity, where the linear potential could be tuned by changing the tilting angle of the tubes. An additional focused laser beam (or other spatial-light-modulator-based techniques) is required to create hard walls at the bottom of the tubes for keeping the atoms confined.

\begin{figure}
\centering
\includegraphics[width=0.48\textwidth]{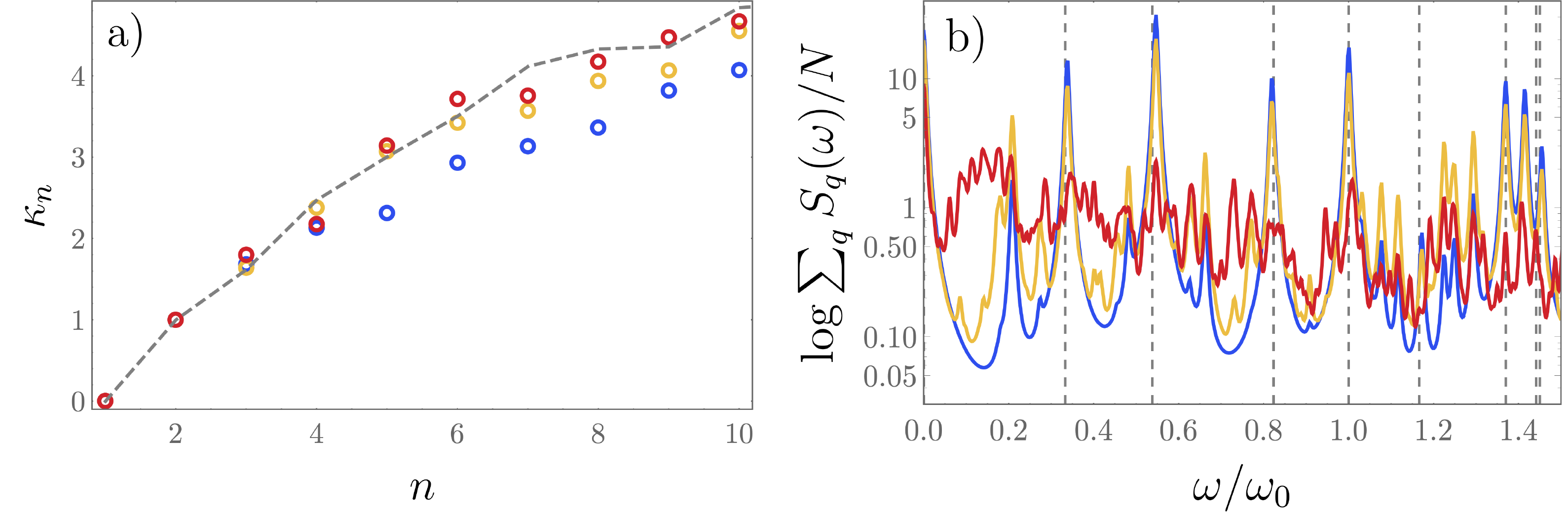}
\caption{a) Comparison of the universal ratios of the Heisenberg model with the ideal coefficients in Eq. \eqref{exactbc} (gray dashed curves) to DMRG simulations of the continuum for fermions in a harmonic trap. Blue, yellow and red circles correspond to interaction strength $g=5,10$ and $15$ (in units of $\hbar^2/ml$), respectively. The universal ratios $\kappa_n$ are calculated having as a reference state $r=2$. b) Temperature-dependent dynamical structure factor (summed over $q$) for harmonically trapped fermions (Eq. \eqref{spinchain}) with $g=25$. Blue, yellow and red curves correspond $k_B T=0.002\epsilon_F$, $0.05 \epsilon_F$ and $0.2\epsilon_F$, respectively, where $\epsilon_F$ is the system's Fermi energy. The vertical gray dashed lines denote the position of the energy gaps corresponding to the universal ratios obtained for the Heisenberg model with couplings given by Eq. \eqref{exactbc}. The reference states for these calculations is $r=4$, and the frequency $\omega_0$ is analogously defined as $\omega_0=(\epsilon_4-\epsilon_0)/\hbar$. In both panels, we assume $N=7$ in a sector of fixed magnetization $+1/2$.}
\label{fig4}
\end{figure}

We also address two other important experimental aspects which can be relevant in the detection of the physical spectrum of trapped systems: finiteness of interactions and temperature. To investigate the first effect, we realize simulations of the continuum with the Hubbard model in an underlying harmonic trap \cite{sm}. In Fig. \ref{fig4} a) we show a comparison of the universal ratios obtained with this approach to the expected results for a Heisenberg model with a set of coefficients given by Eq. \eqref{exactbc}. We find that even at a moderate interaction regime (well within the experimental capabilities) the measured spectrum agrees with the entanglement spectrum of the target model (especially in the low-energy sector at small $n$). We also note that this calculation provides an independent check of the validity of the spin-mapping approach described in the main part of this letter. In experimental setups, interactions between atoms can be manipulated by means of Feshbach \cite{feshbach} or confinement induced resonances \cite{olshanii}.

We quantify the effect of finite temperature on the energy spectrum by calculating the temperature-dependent dynamical structure factor \cite{PhysRevB.79.214408,PhysRevA.91.043617,PhysRevB.97.104424}
\begin{equation}\label{dsf}
S(q,\omega)=\frac{1}{Z(T)}\sum_{i,j}e^{(-E_i/k_B T)}
\vert \langle i \vert S_q^z \vert j \rangle \vert
^2\delta\left[\omega -(E_i-E_j)\right],
\end{equation}
where $Z(T)=\Tr({e^{-H/k_B T}})$ is the canonical partition function, $k_B$ is the Boltzmann constant and $\vert i \rangle$ and $\vert j \rangle$ denote the eigenstates; $S_q^z=\sum_i \sqrt(2)\sin(qi)S_i^z/\sqrt{(N+1)}$ is the Fourier transform of the operator $S_i^z=\sigma_i^z/2$ and $q=n\pi/(N+1)$ are discrete momentum values. In Fig. \ref{fig4} b) we compare the excitation spectrum of the dynamical structure factor (at values of $k_B T$ corresponding to different fractions of the Fermi energy $\epsilon_F$) to the position of the energy gaps corresponding to the universal ratios. Particularly, at lower temperatures we find pronounced peaks located precisely at the energy values predicted by Eq. \eqref{ratios}, that are clearly visible for experimentally-achievable temperatures $0.05\epsilon_F$ (note the vertical logarithmic scale). As expected, for larger temperatures such results are washed out by the contribution of several additional frequencies. We point out that not all excitation peaks present in Fig. \ref{fig4} b) are contemplated with a given definition of the universal ratios given by Eq. \eqref{ratios}.

\paragraph*{Concluding remarks.}
We have presented a theoretical proposal for the realization of entanglement Hamiltonians with strongly interacting cold atoms in effectively one-dimensional optical traps. A key feature of these models is the possibility of mapping the Hamiltonian into a spin chain with a set of couplings that depend on the underlying geometry. Particularly, by simply assuming a harmonic confinement, we have shown that the universal ratios of the entanglement for a partition embedded in an infinite system can be reproduced. Analogously, a linear potential can produce the results expected for a half-partition in a finite system. The energy spectrum of these systems can thus be obtained by standard spectroscopy of cold atoms in elongated harmonic traps or box-like traps under the effect of gravity. We have benchmarked the robustness of our predictions against important experimental effects such as finite interaction strength and finite temperature, evidencing the experimental feasibility of our proposal. Such a protocol can be extended to the study of various spin Hamiltonians, like systems with higher internal symmetries \cite{PhysRevB.94.195110}; in such cases, the particular details (e.g. interactions and internal states) of a given atomic model will determine the structure of the spin chain \cite{xiaoling,laird}. Nevertheless, the relation between the couplings between nearest neighbors and the geometry of the external trap remains valid in the limit of strong interactions.

\paragraph*{Acknowledgements} The authors thank Marcello Dalmonte for discussing the project and for valuable comments on the manuscript and Jacopo Catani for important remarks on the experimental feasibility of the proposal. We acknowledge support from H2020 European Research Council (ERC Consolidator Grant TOPSIM Grant Agreement No. 682629), European QuantERA ERA-NET Cofund in Quantum Technologies (Project QTFLAG Grant Agreement No. 731473), Ministero dell’Istruzione, dell’Università e della Ricerca (MIUR Project FARE TOPSPACE R16SPCCRCW, and MIUR PRIN Project No. 2017E44HRF).
The DMRG calculations shown in this paper were performed using the ITensor library \cite{ITensor}.

\bibliographystyle{ieeetr}
\bibliography{biblio}

\newpage
\widetext

\section*{Supplemental Material}
\subsection*{Mapping a strongly interacting atomic system to a spin chain}
The general procedure described in this section has been developed and extensively detailed in different works, such as \cite{artem1,deuretz2,pu1}.
Further discussions and applications to different atomic models can be found, for instance, in \cite{xiaoling,massignan}. For clarity, we reproduce below the essential steps required to obtain the spin chain models considered in the main text.
We start by considering a one-dimensional two-component fermionic system with contact interaction, described by 
\begin{eqnarray}
H=\sum_{i=1}^{N} H_0(x_i)+g\sum_{i=1}^{N_\uparrow}\sum_{j=1}^{N_\downarrow}\delta(x_{\uparrow i}-x_{\downarrow j})
\end{eqnarray}
where $N=N_\uparrow+N_\downarrow$. In the limit of infinite repulsion ($g\rightarrow \infty$), we can write the complete many-body wave function as
\begin{equation}\label{ansatz}
\Psi=\sum_{k=1}^{L(N_\uparrow,N_\downarrow)}a_k P_k\Phi_0(\{x_{\uparrow i},x_{\downarrow j}\}),
\end{equation} 
where $P_k$ is the permutation operator and the sum is carried over a total number of $L(N_\uparrow,N_\downarrow)=\binom{N_\uparrow+N_\downarrow}{N_\uparrow}$ permutations. In this context $\Phi_0$ is the wave function in the limit of infinite repulsion where $\{x_{\uparrow i},x_{\downarrow j}\}$ denotes a given ordering of the particles.

For strong finite interactions, the Hellmann-Feynman theorem can be employed to write
\begin{eqnarray}\label{hellmann}
\frac{\partial E}{\partial g}&=&\sum_{i=1}^{N_\uparrow}\sum_{j=1}^{N_\downarrow}\langle\Psi\vert\delta (x_{\uparrow i}-x_{\downarrow j})\vert\Psi\rangle
\end{eqnarray}
Additionally, we write the expression for the derivative condition at the contact point between particles as
\begin{equation}\label{cond2}
\left(\frac{\partial\Psi}{\partial x_{\uparrow i}}-\frac{\partial\Psi}{\partial x_{\downarrow j}}\right)\Bigg\rvert_{x_{\uparrow i}-x_{\downarrow j}= 0^-}^{x_{\uparrow i}-x_{\downarrow j}= 0^+}=2g \Psi(x_{\uparrow i}=x_{\downarrow j}),
\end{equation}
which follows the guidelines of the coordinate Bethe ansatz approach \cite{lieb1}. Plugging Eq. \eqref{cond2} into Eq. \eqref{hellmann} and integrating with respect to $g$, we find

\begin{eqnarray}
E=E_0-\frac{\sum_{i=1,j=1}^{N_\uparrow,N_\downarrow}\int dx_{\uparrow 1},\cdots,dx_{\uparrow N_\uparrow}\int dx_{\downarrow 1},\cdots,dx_{\downarrow N_\downarrow}\Bigg\rvert\left(\frac{\partial\Psi}{\partial x_{\uparrow i}}-\frac{\partial\Psi}{\partial x_{\downarrow j}}\right)\bigg\rvert_{x_{\uparrow i}-x_{\downarrow j}= 0^-}^{x_{\uparrow i}-x_{\downarrow j}= 0^+}\Bigg\rvert^2\delta (x_{\uparrow i}-x_{\downarrow j})}{4g\int dx_{\uparrow 1},\cdots,dx_{\uparrow N_\uparrow}\int dx_{\downarrow 1},\cdots,dx_{\downarrow N_\downarrow}\vert \Psi \vert^2}.
\end{eqnarray}
where we discard terms of order $O(1/g^2)$ and higher. In this expression, $E_0$ denotes the energy of a system of spinless fermions (which is the same energy expected in the regime of infinite repulsion). By inserting \eqref{ansatz} in this equation, we find
\begin{eqnarray}\label{functional}
E=E_0-\frac{\sum_{i=1}^{N-1}\frac{\alpha_i}{g}\sum_{k=1}^{L(N_\uparrow-1,N_\downarrow-1)}(a_{ik}-a'_{ik})^2}{\sum_{k=1}^{L(N_\uparrow,N_\downarrow)}a_{k}^2}
\end{eqnarray}
where $a_{ik}$ denotes the wave function coefficient where neighboring $\uparrow$ and $\downarrow$ fermions are found in position $i$ and $i+1$, while $a'_{ik}$ is the coefficient for the wave function where the particle positions are $i+1$ and $i$. In this expression, we have
\begin{equation}\label{geosm}
\alpha_i=\frac{\int_{x_1<x_2\cdots<x_N-1}dx_1...dx_{N-1}\Big|\frac{\partial \Phi_0(x_1,\cdots,x_i,\cdots,x_N)}{\partial x_N}\Big|^2_{x_N=x_i}}{\int_{x_1<x_2\cdots<x_N-1}dx_1\cdots dx_N |\Phi_0(x_1,\cdots,x_i,\cdots,x_N)|^2},
\end{equation}
which is not spin-dependent. Thus, $\Phi_0$ can now be treated as the wave function for a system of spinless fermions.

We now take under consideration a spin Hamiltonian given by
\begin{equation}\label{spinh}
H=E_0-\sum_{i=1}^{N-1}J_i \Pi_{i,i+1},
\end{equation}
where $\Pi_{\uparrow\downarrow}^{i,i+1}=\frac{1}{2}(1-\vec{\sigma}_i\cdot\vec{\sigma}_{i+1})$ is the permutation operator. A general wave function for this Hamiltonian can be written as
\begin{equation}
\vert \chi \rangle=\sum_{k=1}^{L(N_\uparrow,N_\downarrow)}a_k P_k \vert \uparrow_1\cdots \uparrow_{N_\uparrow}\downarrow_1\cdots\downarrow_{N_\downarrow}\rangle,
\end{equation}
in analogy with Eq. \eqref{ansatz}. Using this wave function, we calculate $\langle \chi \vert H\vert \chi\rangle$ as
\begin{eqnarray}\label{functional2}
\langle \chi \vert H\vert \chi\rangle=E_0- 
\frac{\sum_{i=1}^{N-1}J_i\sum_{k=1}^{L(N_\uparrow-1,N_\downarrow-1)}(a_{ik}-a'_{ik})^2}{\sum_{k=1}^{L(N_\uparrow,N_\downarrow)}a_{k}^2}
\end{eqnarray}
where $a_{ik}$ and $a'_{ik}$ are as described for Eq. \eqref{functional}. This means that expressions \ref{functional} and \ref{functional2} are equivalent provided that $J_i=\alpha_i/g$. We can now rewrite Eq.~\ref{spinh} as
\begin{equation}
H=E_0-\frac{1}{2}\sum_{i=1}^{N-1}\frac{\alpha_i}{g} (1-\vec{\sigma}_i\cdot\vec{\sigma}_{i+1}),
\end{equation}
which is the Hamiltonain given in Eq. \eqref{spinchain}.

\subsection*{Details on the simulations}
In Fig. \ref{fig2} a) and b) of the main text, we calculate the spatial distributions for a system of atoms in a harmonic trap and in a linear potential. This quantity is calculated with the following expression:
\begin{equation}\label{onebody}
\rho^i(x)=\int_{\Gamma} dx_1...dx_N \,\delta(x_i-x)|\Phi_0(x_1,...,x_i,...,x_N)|^2,
\end{equation} 
where the integration is restricted to the sector $\Gamma=x_1 <x_2<...<x_N$. For larger systems, it is convenient to explore the determinant properties of the wave function $\Phi_0$ \cite{deuretz1}, and write
\begin{equation}\label{dets}
\rho^i(x)=\frac{\partial}{\partial x}\left( \sum_{j=0}^{N-1}\frac{(-1)^{N-1}(N-j-1)!}{(i-1)!(N-j-i)!j!} \frac{\partial^j}{\partial \lambda^j}\det \left[B(x)-1\lambda \right]\vert_{\lambda=0}\right),
\end{equation}
where the elements of the matrix $B(x)$ are written as $b_{mn}(x)=\int_{-\infty}^{x}dy\,\varphi_m(y)\varphi_n(y)$, and $\varphi(x)$ denotes the single-particle states in a corresponding trapping potential. For the simple case of the harmonic trap, these can be obtained exactly; for the linear potential, we obtain the single particle solutions by numerical diagonalization, using as a basis 50 eigenstates of the box potential. The characteristic length $l$ in the harmonic trap is related to the trapping frequency $\omega$ by $l=\sqrt{\hbar/m\omega}$ (we assume $\omega=1$ in our calculations).
The exchange coefficients shown in Fig. \ref{fig2} c) and d) can be calculated for small systems with Eq. \eqref{geo}. For larger system, we use the open-source program CONAN \cite{conan}.
Most results found in Fig. \ref{fig3} are obtained by considering the full energy spectrum calculated by exactly diagonalizing the corresponding Hamiltonians. In c) and d), the black curves are calculated with DMRG simulations of the Heisenberg model, with a total of 6 DMRG sweeps and a maximum bond dimension of 600. 

In Fig. \ref{fig4}, we obtain the results at intermediate interactions in a system of $N=7$ fermions with DMRG by approximating the continuum with the Hubbard model

\begin{equation}
H=-t\sum_{j,\sigma}(c^\dagger_{j+1,\sigma}c_{j,\sigma}+\text{H.c.})+U\sum_{j}n_{j,\uparrow}n_{j,\downarrow}+\sum_{j,\sigma}V_{j}n_{j,\sigma},
\end{equation}
where the last term denotes the underlying harmonic trap potential. The simulation of the continuum is performed with a total of $N_S=200$ sites. By fixing a length $\lambda$ we define the lattice spacing as $a=\lambda/N_s$. The hopping parameter is then related to the kinetic term in the continuum as $t=1/(2a^2)$ (assuming $m=1$), while the interaction parameters are related by $U=g/a$ \cite{kleine}. In these simulations we perform a total of 40 DMRG sweeps, with a maximum bond dimension of $10^6$. 

\subsection*{Additional results}

\begin{figure}
\centering
\includegraphics[width=0.8\textwidth]{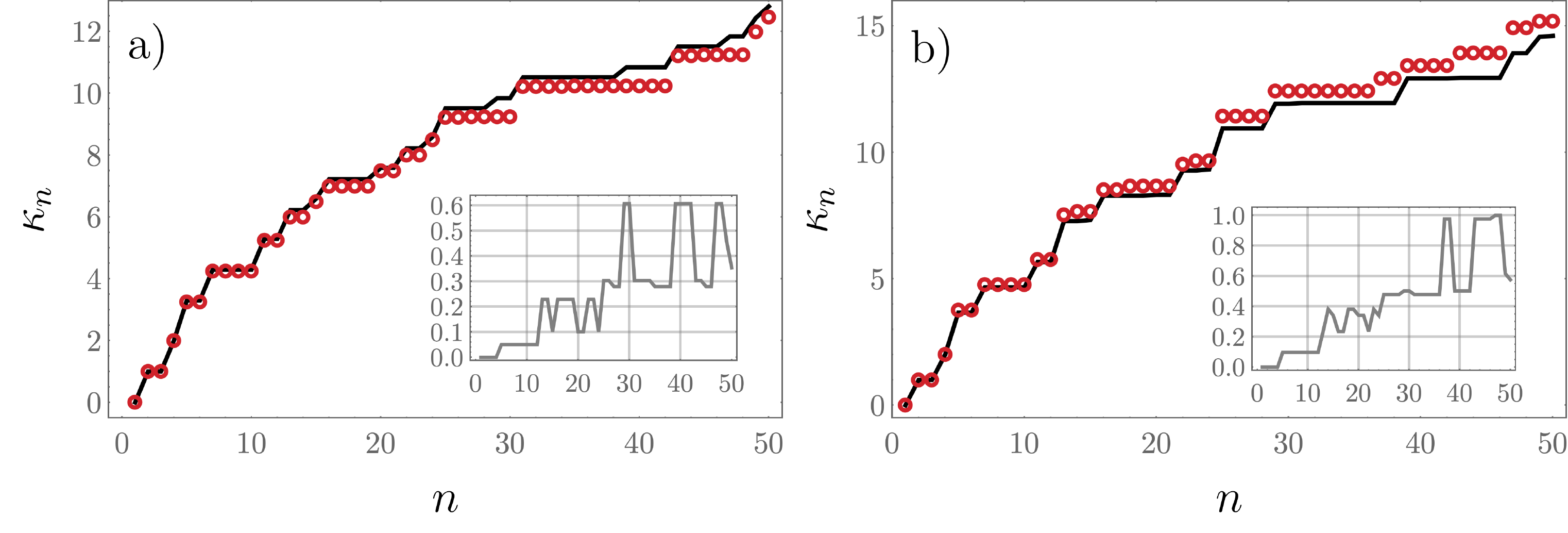}
\caption{Comparison of the universal ratios in the XX model for systems with $N=10$ in the case of a) the harmonic trap and b) the linear potential. The results for the partitions with a) PBC conditions and b) OBC (black curves) are obtained with DMRG for systems of total size $L=80$ and $L=20$, respectively. The interaction parameters for the trapped atomic systemare set as $g=20$ and $\kappa=2$ (see main text).}
\label{figsm1}
\end{figure}

In the main text we showed, in Fig. \ref{fig3} e)-f), the comparison between the universal ratios obtained with ED for trapped atoms and DMRG for partitions embedded in larger systems, focusing on the Heisenberg model. Here we present in Fig. \ref{figsm1} also the case of the XX model, where the size of the system is the same as considered in the Fig. \ref{fig3} e)-f). As in the main text, we observe an increasing discrepancy for higher energy levels, which can be partially explained by a lack of convergence of the DMRG runs for the choice of parameters described above. This results can be therefore improved by considering additional runs with larger bond dimensions.

\begin{figure}
\centering
\includegraphics[width=0.8\textwidth]{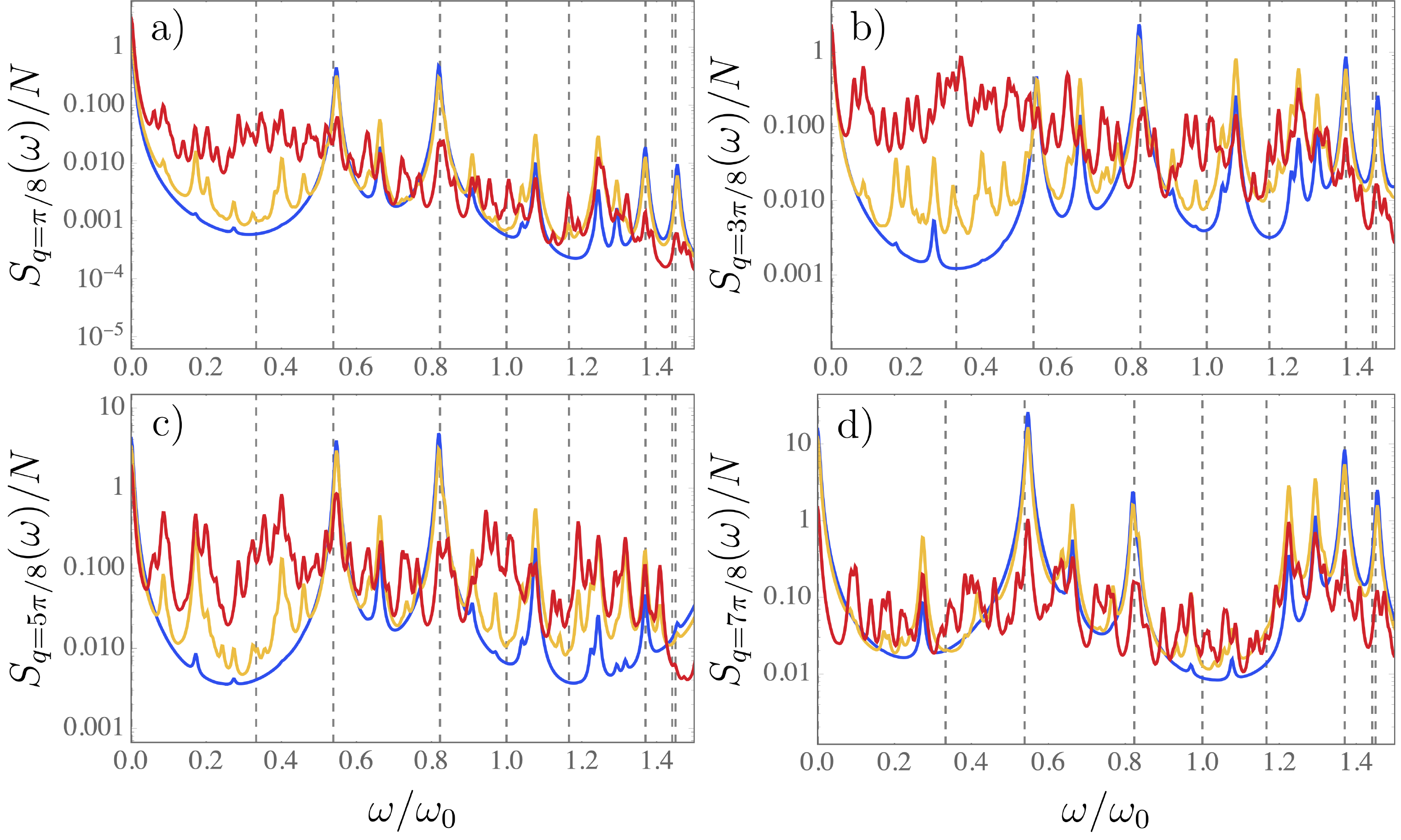}
\caption{Temperature-dependent dynamical structure factor for harmonically trapped fermions (Eq. \eqref{spinchain}) with $g=25$, in the cases of a) $q=\pi/8$, b) $q=3\pi/8$, c) $5\pi/8$, d) $7\pi/8$. Blue, yellow and red curves correspond $k_B T=0.002\epsilon_F$, $0.05 \epsilon_F$ and $0.2\epsilon_F$, respectively, where $\epsilon_F$ is the system's Fermi energy. The vertical gray dashed lines denote the position of the energy gaps corresponding to the universal ratios obtained for the Heisenberg model with couplings given by Eq. \eqref{exactbc}. The reference states for these calculations is $r=4$, and the frequency $\omega_0$ is analogously defined as $\omega_0=(\epsilon_4-\epsilon_0)/\hbar$. We assume $N=7$ in a sector of fixed magnetization $+1/2$.}
\label{figsm2}
\end{figure}

In the results contained in the main text for the temperature-dependent dynamical structure factor, we approximate the delta function $\delta\left[\omega -(E_i-E_j)\right]$ contained in Eq. \eqref{dsf} with a Lorentzian given by
\begin{equation}
    f(\omega)=\frac{1}{\pi}\frac{\eta^2}{\eta^2+\omega^2}
\end{equation}
where $\eta=0.002$. In Fig. \ref{fig4} b) we summed over all values of momentum $q$. In Fig. \ref{figsm2} we also show the separate results for this quantity at particular momentum values.

\end{document}